\documentstyle[sprocl,psfig]{article}

\def\be{\begin{equation}}
\def\ee{\end{equation}}
\def\bea{\begin{eqnarray}}
\def\eea{\end{eqnarray}}

\begin{document}

\begin{flushright}
\begin{tabular}{l}
CERN-TH/98-218 \\
hep-ph/9806529
\end{tabular}
\end{flushright}
 
\vspace{8mm}

\begin{center}
 
{\Large \bf Supersymmetric Q-balls: theory and cosmology}\footnote{Talk
  presented at the Sixth International Symposium on Particles, Strings and
  Cosmology ({\bf PASCOS-98}), Northeastern University, March 22-29, 1998.}

\vspace{8mm}

{\large
Alexander Kusenko}\footnote{ email address: Alexander.Kusenko@cern.ch}
 
\vspace{6mm}
Theory Division, CERN, CH-1211 Geneva 23, Switzerland \\

\vspace{12mm} 
 
{\bf Abstract}

\end{center}

MSSM predicts the existence of Q-balls, some of which
can be entirely stable.  Both stable and unstable Q-balls can play an
important role in cosmology.  In particular, Affleck--Dine baryogenesis
can  result in a copious production of stable baryonic Q-balls, which
can presently exist as a form of dark matter.
 
\vspace{40mm}
 
\begin{flushleft}
\begin{tabular}{l}
CERN-TH/98-218 \\
June, 1998
\end{tabular}
\end{flushleft}

\vfill
 
\pagestyle{empty}
 
\pagebreak
 
\pagestyle{plain}
\pagenumbering{arabic}

\title{Supersymmetric Q-balls: theory and cosmology
}

\author{Alexander Kusenko}

\address{Theory Division, CERN, CH-1211 Geneva-23, Switzerland
\\E-mail: Alexander.Kusenko@cern.ch} 


\maketitle\abstracts{ MSSM predicts the existence of Q-balls, some of which
  can be entirely stable.  Both stable and unstable Q-balls can play an
  important role in cosmology.  In particular, Affleck--Dine baryogenesis
  can  result in a copious production of stable baryonic Q-balls, which
  can presently exist as a form of dark matter.  
}

In a class of theories with interacting scalar fields $\phi$ that carry
some conserved global charge, the ground state is a Q-ball~\cite{q}, a lump
of coherent scalar condensate that can be described semiclassically as a 
non-topological soliton and has a form 
\begin{equation}
\phi(x,t) = e^{i \omega t} \bar{\phi}(x).
\label{q}
\end{equation}
Q-balls exist whenever the scalar potential satisfies certain conditions
that were first derived for a single charged degree of freedom~\cite{q} and
were later generalized to a theory of many scalar fields with different
charges~\cite{ak_mssm}.     

It was recently pointed out that all phenomenologically viable
supersymmetric extensions of the Standard Model predict the existence of
non-topological solitons~\cite{ak_mssm} associated with the conservation of
baryon and lepton number.  The MSSM admits a large number of different
Q-balls, characterized by (i) the quantum numbers of the fields that form
a spatially-inhomogeneous ground state and (ii) the net global charge of
this state.   

First, there is a class of Q-balls associated with the
tri-linear interactions that are inevitably present in the
MSSM~\cite{ak_mssm}.  The masses of such Q-balls grow linearly with their
global charge, which can be an arbitrary integer number~\cite{ak_qb}.
Baryonic and leptonic Q-balls of this variety are, in general, unstable
with respect to their decay into fermions.  However, they could form in the
early universe through the accretion of global
charge. In the false vacuum such a process could precipitate an otherwise
impossible or slow phase transition~\cite{ak_pt}.  In a metastable vacuum,
a  Q-ball tends to have a negative energy density in its interior.  When
the charge reaches some  critical value, a Q-ball expands and converts
space into a true-vacuum phase.  In the case of tunneling, the critical
bubble is formed through coincidental coalescence of random quanta into an
extended coherent  object.  This is a small-probability event.  If,
however, a Q-ball grows through charge accretion, it reaches the critical
size with probability one, as long as the conditions for
growth~\cite{ak_pt} are satisfied.  Therefore, the phase transition can
proceed at a much faster rate than it would by tunneling.   

The second class~\cite{dks} of solitons comprises the Q-balls whose VEV is
aligned with some flat direction of the MSSM and is a
gauge-singlet~\cite{kst} combination of squarks and sleptons with a 
non-zero baryon or lepton number.  The potential along a flat
direction is lifted by some soft supersymmetry-breaking terms that originate
in a ``hidden sector'' of the theory some scale $\Lambda_{_S}$ and are
communicated to the observable sector by some interaction with a coupling
$g$, so that $g \Lambda \sim 100$~GeV.  Depending on the strength of the
mediating interaction, the scale $\Lambda_{_S}$ can be as low as a few TeV 
(as in the case of gauge-mediate SUSY breaking), or it can be  some
intermediate scale if the mediating interaction is weaker (for instance, 
$g\sim \Lambda_{_S}/m_{_{Planck}}$ and $\Lambda_{_S}\sim 10^{10}$~GeV in
the case of gravity-mediated SUSY breaking).  For the lack of a definitive
scenario, we take $\Lambda_{_S}$ to be a free parameter.   Below
$\Lambda_{_S}$ the mass terms are generated for all the scalar degrees of
freedom, including those that parameterize the flat direction.  At the
energy scales larger than $\Lambda_{_S}$, the mass terms turn off and the
potential is ``flat'' up to some logarithmic corrections.  If the Q-ball 
VEV extends beyond $\Lambda_{_S}$, the mass of a soliton~\cite{dks,ks} is
no longer proportional to its global charge $Q$, but rather to $Q^{3/4}$. 

This allows for the existence of some entirely stable Q-balls with a large
baryon number $B$ (B-balls).  Indeed, if the mass of a B-ball is $M_{_B} \sim
({\rm 1~TeV}) \times B^{3/4}$, then the energy per baryon number
$(M_{_B}/B)\sim ({\rm 1~TeV}) \times B^{-1/4}$ is less than 1~GeV for $B >
10^{12}$.  Such large B-balls cannot  dissociate into protons and neutrons
and are entirely stable thanks to the conservation of energy and the baryon
number.  If they were  produced in the early universe, they would exist at
present as a form of dark matter~\cite{ks}.  

At the end of inflation, the scalar fields of the MSSM develop some large
expectation values along the flat directions, some of which have
a non-zero baryon number~\cite{ad}. Initially, the scalar condensate has the
form given in eq.~(\ref{q}) with $\bar{\phi}(x)= const$ over the length
scales greater than a horizon size. One can think of it as a universe
filled with Q-matter.  The relaxation of this condensate to the potential
minimum is the cornerstone of the Affleck--Dine (AD) scenario for
baryogenesis.  

It was often assumed that the condensate remains spatially homogeneous from
the time of formation until its decay into the matter baryons.  This
assumption is, in general, incorrect.  In fact, the initially homogeneous
condensate can become unstable~\cite{ks} and break up into Q-balls whose
size is determined by the potential and the rate of expansion of the
Universe.  B-balls with $12 < \log_{10} B < 30$ can form naturally
from the breakdown of the AD condensate.  These are entirely
stable if the flat direction is ``sufficiently flat'', that is if the
potential grows slower than $\phi^2$ on the scales or the order of
$\bar{\phi}(0)$.   The evolution of the primordial condensate can be
summarized as follows: 

\vspace{3mm}
\psfig{figure=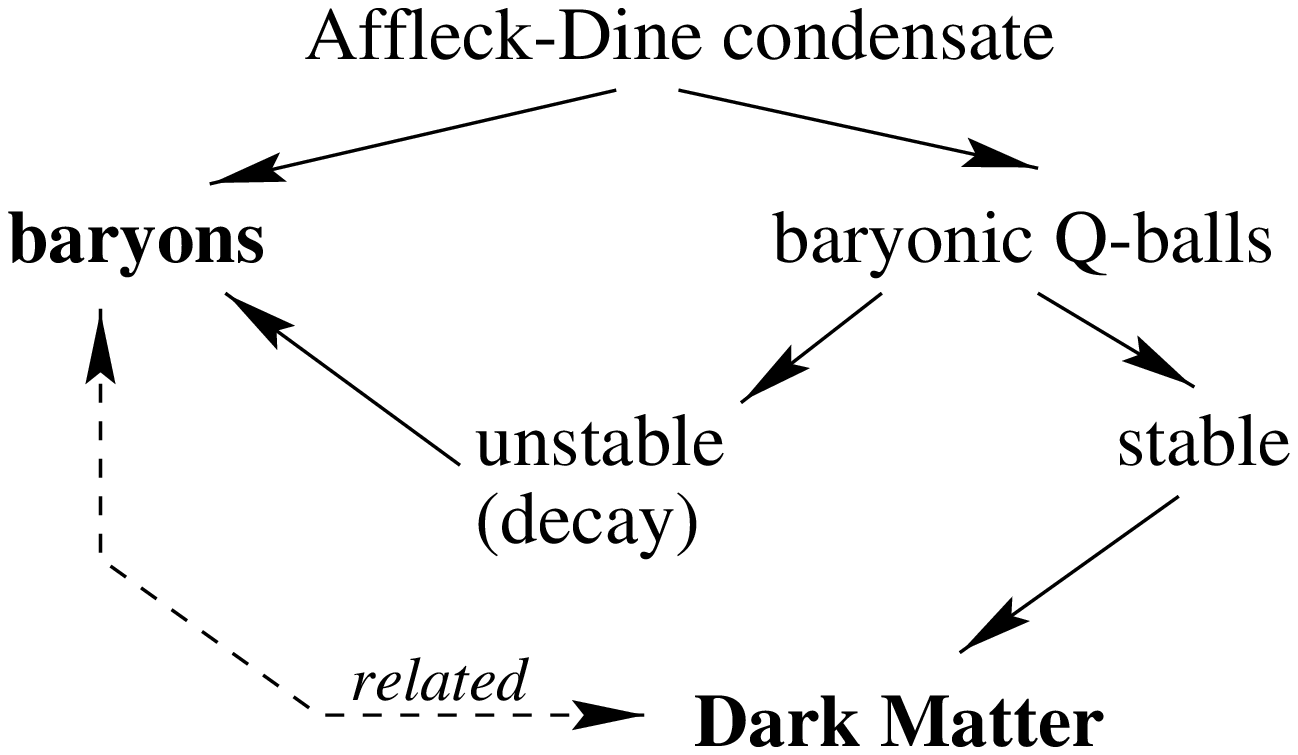,height=1.5in,width=3.5in}

Conceivably, the cold dark matter in the Universe can be made up entirely 
of SUSY Q-balls.  Since the baryonic matter and the dark matter share the
same origin in this scenario, their contributions to the mass density of
the Universe are related.  Therefore, it is easy to understand why the
observations find $\Omega_{_{DARK}} \sim \Omega_{B} $ within
an order of magnitude.  This fact is extremely difficult to explain in
models that invoke a dark-matter  candidate whose present-day abundance is
determined by the process of freeze-out, independent of baryogenesis.  If
this is the case, one could expect $\Omega_{_{DARK}}$ and $\Omega_{B} $ to
be different by many orders of magnitude.  If one doesn't want to accept
this equality as fortuitous, one is forced to hypothesize some {\it ad hoc} 
symmetries~\cite{kaplan} that could relate the two quantities.   In the 
MSSM with AD baryogenesis, the amounts of dark-matter Q-balls and the
ordinary matter baryons are naturally related~\cite{ks}.  One
predicts~\cite{lsh} $\Omega_{_{DARK}} = \Omega_{B} $  for B-balls with  $B
\sim 10^{26}$. 

This size is well above the present experimental lower limit on the baryon
number of an average relic B-ball, under the assumption that all or most of
cold dark matter is made up of Q-balls.  On their passage through matter,
the electrically neutral baryonic SUSY Q-balls can cause a proton decay,
while the electrically charged B-balls produce massive ionization.
Although the condensate inside a Q-ball is electrically neutral~\cite{kst},
it may pick up some electric charge through its interaction with
matter~\cite{kkst}.  Regardless of its ability to retain electric charge,
the Q-ball would produce a straight track in a detector and would release
the energy of, roughly, 10 GeV/mm.  The present limits~\cite{kkst}
constrain the baryon number of a relic dark-matter B-ball to be greater
than $10^{22}$.  Future experiments are expected to improve this limit.
It would take a detector with the area of several square kilometers to
cover the entire interesting range $B\sim 10^{22} ... 10^{30}$.  

The relic Q-balls can accumulate in neutron stars and can lead to their
ultimate destruction over a time period from one billion years to longer
than the age of the Universe~\cite{sw}.  If the lifetime of a neutron star
is in a few Gyr range, the predicted mini-supernova explosions may be
observable and may be related to the gamma-ray bursts. 

A different scenario that relates the amounts of baryonic and dark matter
in the Universe, and in which the dark-matter particles are produced from
the decay of unstable B-balls was proposed by Enqvist and
McDonald~\cite{em}. 


The following diagram illustrates our conclusions and the assumptions on
which they rely: 

\vspace{3mm} 

\psfig{figure=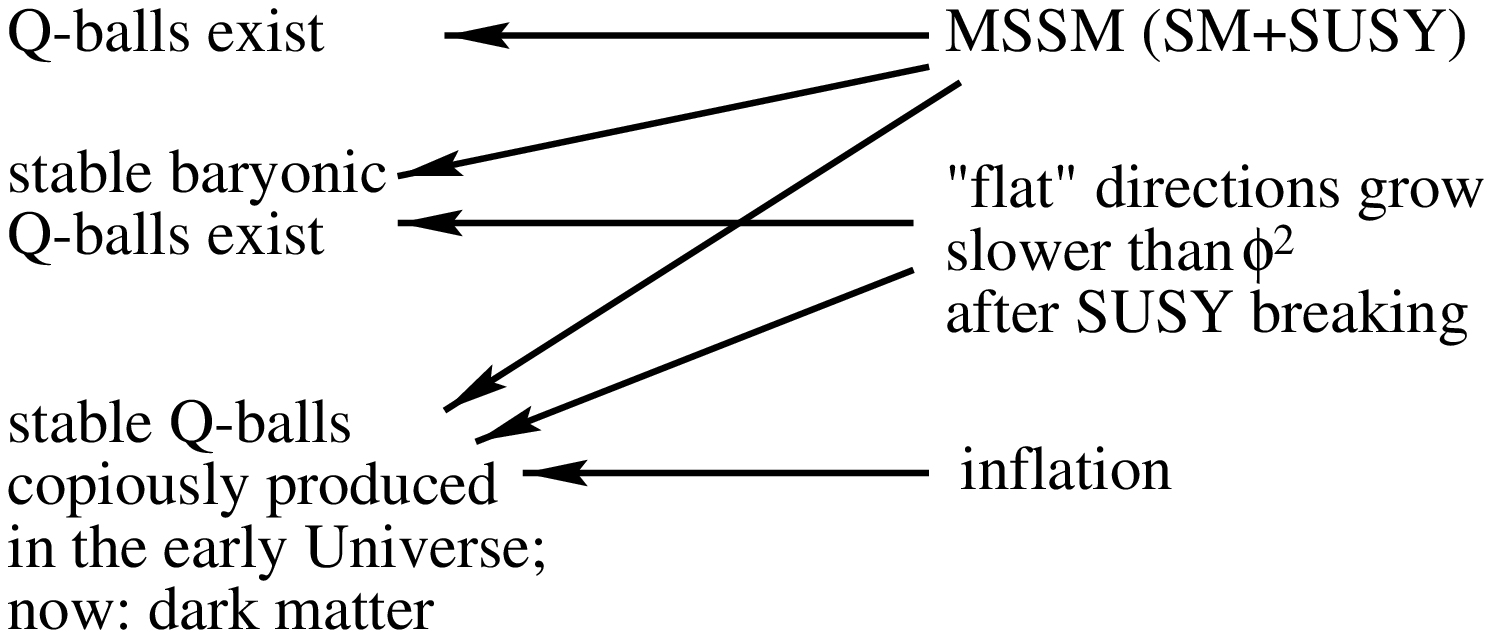,height=1.4in,width=3.8in}

\end{document}